\documentstyle[12pt]{article}
\begin{document}
\rightline{YCTP-P16-92}
\rightline{May 1992}
\vskip .5in
\begin{center}
{\bf\large Contribution of Long Wavelength Gravitational Waves to the Cosmic
Microwave Background Anisotropy}
\end{center}
\vskip .1in
\begin{center}
Martin White\footnote{Address after September 1, Center for Particle
Astrophysics, University of California, Berkeley}

{\it    Center for Theoretical Physics, Sloane Laboratory }

{\it    Yale University, New Haven CT 06511}
\end{center}

\vskip .5in

\centerline{ {\bf Abstract} }

\noindent
We present an in depth discussion of the production of gravitational waves from
an inflationary phase that could have occurred in the early universe, giving
derivations for the resulting spectrum and energy density.
We also consider the large-scale anisotropy in the cosmic microwave background
radiation coming from these waves.
Assuming that the observed quadrupole anisotropy comes mostly from
gravitational waves (consistent with the predictions of a flat spectrum of
scalar density perturbations and the measured dipole anisotropy) we describe in
detail how to derive a value for the scale of inflation of
$(1.5-5)\times 10^{16}$GeV, which is at a particularly interesting scale for
particle physics.  This upper limit corresponds to a 95\% confidence level
upper limit on the scale of inflation assuming only that the quadrupole
anisotropy from gravitational waves is not cancelled by another source.
Direct detection of gravitational waves produced by inflation near this scale
will have to wait for the next generation of detectors.

\vskip .4in

\centerline{To appear in: {\em Physical Review D} }

\newpage

\noindent 1. {\bf Introduction}

It has long been realized that a period of inflation in the early universe
would lead to production of a well defined spectrum of gravitational waves
\cite{Star} (while inflation is not the only method of generating a stochastic
background of horizon-sized waves \cite{LMK}, it is certainly the most well
motivated).
Since the waves decouple early they are a potentially good probe
of conditions in the early universe.  In fact as pointed out in
\cite{Anis,FabPol,AbbWis1,AbbSch} such waves, which are fluctuations in the
metric, can result in distortions of the cosmic microwave background radiation
(CMBR) thus allowing information on the amplitude of the waves (and the
parameters of the inflationary phase that produced them) to be derived from
the measured CMBR anisotropy.
This program proceeds in 3 steps.  First the expected fluctuation spectrum is
derived, then the consequences of such a spectrum for CMBR anisotropies are
calculated.  Finally comparing these predictions with the observations allows
us to infer the parameters of an inflationary theory that produces such waves.

Recently an analysis of the implications of a CMBR anisotropy connected with
gravitational waves from inflation was presented \cite{KraWhi}.
In this paper we fill in details of the results presented in \cite{KraWhi}
and further examine the issues addressed therein.
A review of each of the 3 steps mentioned above will be given (many of the
analytic results presented here have appeared in some form scattered in the
literature, we have attempted to check, unify and reconcile previous
estimates).
The results can be compared to the recent COBE data \cite{COBE} which gives
the first positive measurement of anisotropies in the CMBR (previously this
anisotropy had only been limited from above \cite{COBEold,oldlim}).
On the theoretical side we give the details of derivations of some of the key
results in the hope that these will be of use.  We present in one place the
derivation of the scale invariant spectrum predicted by exponential inflation
(and a description of the scalar analogy), the energy density $\rho$ or
equivalently $\Omega_g$, the CMBR anisotropy produced and the predicted
temperature correlation function.
We show how ``redshift during horizon crossing" dramatically reduces the
predicted $\Omega_g$ from their asymptotic value for waves re-entering the
horizon during the matter dominated epoch.
Our results for the multipole moments of the CMBR temperature fluctuation agree
with \cite{AbbWis1} (though we use an approach due to \cite{AbbHar} to derive
the scale invariant spectrum) and are a factor of $\approx2$ larger than
\cite{FabPol}.  With these results in hand we describe how the recent COBE
observation is consistent with inflation and can limit the ``scale near the
end of inflation" assuming that the CMBR anisotropy is due, at least in part,
to inflation-produced gravitational waves (the contribution from scalar density
perturbations is constrained by the dipole anisotropy as we shall discuss).
The scale turns out to be few$\times 10^{16}$GeV \cite{KraWhi} -- an
interesting energy for particle physics.

The outline of the paper is as follows:  section 2 establishes the
notation and conventions.  In sections 3 and 4 we discuss the scale-invariant
spectrum predicted by exponential inflation and the spectral energy density.
Section 5 is devoted to the derivation of the anisotropy generated by a
stochastic spectrum of gravitational waves.   The relation to the temperature
correlation function measured by double- and triple-beam experiments (e.g.
COBE) is derived in section 6 and the current measurement of the CMB
anisotropy is compared to the theoretical predictions in section 7.
Section 8 contains the conclusions.

\vspace{.1in}

\noindent 2. {\bf Cosmological history}

We will be interested in a cosmology which had an early period of
inflation (where the dominant energy density resided in the vacuum).  To be
specific let us suppose that the universe underwent exponential inflation,
then became radiation dominated and is currently in a matter dominated phase.
The assumption of exponential inflation is a good approximation for
``successful" inflation models \cite{SteTur} based on general relativity since
the requirement of a ``slow roll" period in the evolution of the inflaton
field requires a very flat potential.  The other possibility is that of power
law inflation, where the scale factor grows like a power of time.  This
changes the energy density spectrum (putting more power in long wavelength
modes) and enhances the CMBR anisotropies for small $l$ \cite{AbbWis1}.
It seems difficult to have consistent power law inflation even in Brans-Dicke
theory and we will not consider it further.

Our metric is of the usual $k=0$ Robertson-Walker form
\begin{equation}
  ds^2 = -dt^2 + R^2(t) \left( dx^2 + dy^2 + dz^2 \right) =
  R^2(\tau)\left( -d\tau^2 + d\vec{x}^2 \right)
\label{eqn:metric}
\end{equation}
where $d\tau = dt/R(t)$ is the conformal time.
For an equation of state $p=q\rho$ the scale factor is a power of
the conformal time, $R(\tau)\sim \tau^{2/(1+3q)}$.
For later convenience we list below the scale factor,
which we normalize to unity today, for the history described
above along with the dominant form of energy density for each epoch.
We assume that the transitions are sudden and match $R(\tau)$ and
$\dot{R}(\tau)$ at the transition points (this will be a good enough
approximation for our purposes) \cite{AbbHar}.
This matching implies that $\tau$ is discontinuous across each transition and
$R(\tau)$ has the form
\begin{equation}
R(\tau) = \left\{
\begin{array}{clrc}
-(H \tau)^{-1}       & \mbox{vacuum}    & \tau\in & (-\infty,-\tau_2) \\
2\tau_1\tau/\tau_0^2 & \mbox{radiation} & \tau\in & (\tau_2,\tau_1/2) \\
\tau^2/\tau_0^2      & \mbox{matter}    & \tau\in & (\tau_1,\tau_0)
\end{array} \right.
\label{eqn:scalef}
\end{equation}
where $H$ is the Hubble constant during inflation.  Equating
$\int_0^{\tau_x} d\tau$ with $\int_0^{t_x} dt/R(t)$ we find $\tau_0=3t_0$ is
the conformal time today and $\tau_1=(36t_1 t_0^2)^{1/3}$ is the conformal
time at matter-radiation equality (both $\tau_0$ and $\tau_1$ refer to $\tau$
as measured in the matter dominated era -- in this respect our notation differs
from that in \cite{AbbHar}).  The conformal time at the end of inflation is
$\tau_2 = \tau_0/\sqrt{2H\tau_1}$.
The Hubble constant $H$ and vacuum energy density $V_0$ driving the
inflation are related by
\begin{equation}
  H^2 = {8\pi\over 3} {V_0\over m_{Pl}^2} = {8\pi\over 3} m_{Pl}^2 v
\label{eqn:hubble-v0}
\end{equation}
where $v\equiv V_0/m_{Pl}^4$.

It is useful to have an expression for the size of a wave, relative
to the horizon size.  During a normal FRW expansion the coordinate radius of
the horizon grows as $dr = dt/R(t) = d\tau$.  If the period of inflation is
long the coordinate radius of the horizon is very nearly zero at the beginning
of the radiation dominated era.
Thus at time $t_{*}$ the proper radius of the horizon is
$\approx R(\tau_{*})\tau_{*}$.
If we consider a wave just entering the horizon at this time we find its
comoving wavenumber, $k$, must satisfy
\begin{equation}
  \lambda_{phys} = R_{*}\tau_{*} \Rightarrow k \equiv
  {2\pi R\over\lambda_{phys}} = {2\pi\over\tau_{*}}
\end{equation}
This leads us to the following two limits.  A wave with $k\tau \ll 2\pi$ is
well outside the horizon while a wave with $k\tau \gg 2\pi$ is well within
the horizon.

\vspace{.1in}

\noindent 3. {\bf The scalar analogy}

A classical gravitational wave in the linearized theory is a ripple on the
background space-time
\begin{equation}
  g_{\mu\nu} = R^2(\tau) \left( \eta_{\mu\nu} + h_{\mu\nu} \right)
  \ \ \mbox{where }\ \eta_{\mu\nu}=\mbox{diag}(-1,1,1,1),
  \ \ h_{\mu\nu} \ll 1
\label{eqn:wave-metric}
\end{equation}
In what follows we will work in transverse traceless (TT) gauge and denote
the two independent polarization states of the wave as $+,\times$.  In the
linear theory the TT metric fluctuations are gauge invariant (they can be
related to components of the curvature tensor) \cite{MTW}.
We can write a plane wave with comoving wavenumber $|\vec{k}|$
\begin{equation}
  h_{\mu\nu}(\tau,\vec{x}) = h_{\lambda}(\tau;\vec{k})
  e^{i\vec{k}\cdot\vec{x}} \epsilon_{\mu\nu}(\vec{k};\lambda)
\label{eqn:plane-wave}
\end{equation}
where $\epsilon_{\mu\nu}(\vec{k};\lambda)$ is the polarization tensor and
$\lambda=+,\times$.  The equation for the amplitude
$h_{\lambda}(\tau;\vec{k})$ is obtained by requiring the perturbed metric
(\ref{eqn:wave-metric}) satisfy Einstein's equations to $O(h)$.  One finds
\cite{MTW}
\begin{equation}
  \ddot{h}_{\lambda}+2{\dot{R}\over R}\dot{h}_{\lambda}+k^2 h_{\lambda} = 0
\label{eqn:wave-evol}
\end{equation}
As noticed by Grishchuk \cite{Grishchuk} this is just the massless Klein-Gordon
equation for a plane wave in the background space-time.  Thus each
polarization state of the wave behaves as a massless, minimally coupled, real
scalar field, with a normalization factor of $\sqrt{16\pi G}$ relating the two.
We can use the following heuristic argument to fix the $\sqrt{16\pi G}$:
in the linearized theory one can write the Hilbert action for the
fluctuation using the contribution to the Ricci scalar from $h_{\mu\nu}$.
{}From the derivation of the stress-energy tensor $T_{\mu\nu}^{(h)}$ \cite{MTW}
one sees $R^{(h)}={1\over 4}h_{\mu\nu;\rho}h^{\mu\nu;\rho}$ so the
corresponding Hilbert action is
\begin{equation}
  S_H^{(h)} = {\sqrt{-g}\over 16\pi G}\ {1\over 2}
  \left[ (\nabla h_{+})^2 + (\nabla h_{\times})^2 \right]
\end{equation}
which is the action for two real, massless, scalar fields
$\phi_{+,\times} = (16\pi G)^{-1/2} h_{+,\times}$ as expected.
Thus the study of quantum mechanical graviton production reduces to the study
of the fluctuations of a scalar field in the curved background space-time, a
factor $\sqrt{16\pi G}$ relating the two cases.

\vspace{.1in}

\noindent 4. {\bf The spectrum and $\Omega_g$}

Quantum fluctuations have an important consequence in a cosmology with
inflation.  During inflation short wavelength quantum fluctuations will get
red-shifted out of the horizon, after which they freeze in.
In terms of (\ref{eqn:wave-evol}) the freezing in of extrahorizontal modes is
the statement that when wavelength is much larger than the horizon the $k^2 h$
term is negligible and the solution is a constant amplitude: $\dot{h}=0$
(even before this limit is reached we note that the behaviour of waves outside
the horizon is qualitatively different from the damped oscillatory behaviour
of waves within it).
When the mode re-enters the horizon at a much later epoch it appears as a long
wavelength, classical gravitational wave (in analogy with the case of scalar
fluctuations considered by \cite{GuthPi}).  One way of thinking of this
\cite{Krauss} is that the number of quanta describing a state of constant
amplitude grows as a power of the scale factor, the energy of each quantum
being redshifted.  Thus when the mode re-enters the horizon it represents a
very large number of quanta.

The spectrum of gravitational waves generated by quantum fluctuations during
the inflationary period can be derived by a sequence of Bogoliubov
transformations relating creation and annihilation operators defined in the
various phases: inflationary, radiation and matter dominated
\cite{AbbHar,Sahni}.  The key idea is that for modes which have inflated
outside the horizon the transitions between the phases are sudden and the
universe will remain in the quantum state it occupied before the transition
(treating each of the transitions as instantaneous is a good approximation for
all but the highest frequency graviton modes).  However the creation and
annihilation operators that describe the particles in the state are related
by a Bogoliubov transformation, so the quantum expectation value of any
string of fields is changed (see e.g. \cite{Brand} for a discussion of
the calculation of Bogoliubov coefficients).

We calculate the statistical average of the ensemble of classical waves by
considering the corresponding quantum average.  The simplest examples to
consider are the 2-point functions.  For the quantum theory we calculate the
scalar 2-point function, following \cite{AbbHar}
\begin{equation}
\Delta \equiv {k^3\over (2\pi)^3} \int d^3x\ e^{i\vec{k}\cdot\vec{x}}
  \left\langle\psi \right| \phi(\vec{x},\tau)\phi(\vec{0},\tau)
  \left| \psi \right\rangle \ \ \ \mbox{ where }\
  \left|\psi\right\rangle = \left|\mbox{de Sitter vac.}\right\rangle
\label{eqn:delta-defn}
\end{equation}
with the field
\begin{equation}
\phi\left(\vec{x},\tau\right) = \int d^3k \left(
  e^{ i\vec{k}\cdot\vec{x}} a(k)\phi_k(\tau) +
  e^{-i\vec{k}\cdot\vec{x}} a^{\dagger}(k)\phi_k^{*}(\tau) \right)
\end{equation}
where $\phi_k(\tau)$ is a properly normalized solution of
(\ref{eqn:wave-evol}).  For the case of interest, a matter dominated universe
today, $\phi_k$ is related to a Hankel function of order $3/2$
\begin{equation}
  \phi_k(\tau)={e^{-ik\tau}\over\sqrt{2k}\ (2\pi)^{3/2}\ R(\tau)}
    \left(1-{i\over k\tau}\right)
\label{eqn:norm-soln}
\end{equation}
In the exponentially inflating phase $\phi_k$ has the same functional form
(with different $R(\tau)$) and in the radiation dominated phase the
form is again as (\ref{eqn:norm-soln}), with appropriate $R(\tau)$ and
also with the factor in parentheses absent.

The quantum 2-point function is obtained by using the Bogoliubov coefficients
relating the creation and annihilation operators $a$ and $a^{\dagger}$ of the
field in the 3 phases
\begin{eqnarray}
  a_{rad} & = & c_1(\vec{k}) a_{inf}(\vec{k}) +
    c_2^{*}(\vec{k})a^{\dagger}_{inf}(-\vec{k}) \\
  a_{mat} & = & c_3(\vec{k}) a_{inf}(\vec{k}) +
    c_4^{*}(\vec{k})a^{\dagger}_{inf}(-\vec{k}) \label{eqn:bog-mat-inf}
\end{eqnarray}
We are interested in waves which are still well outside the horizon at the time
of matter-radiation equality $(k\tau_1\ll 2\pi)$ since these will give the
largest contribution to the CMBR anisotropy today.  Matching the field and its
first derivative at $\tau_2,\tau_1$ in the limit $k\tau\ll 2\pi$ we find
\begin{equation}
c_1 \approx -c_2 \approx {-H\tau_1\over (k\tau_0)^2}
\ \ \ \ , \ \ \ \ c_3 \approx c_4^{*} \approx {-3iH\over 2k^3\tau_0^2}
\label{eqn:bog-coeff}
\end{equation}
which is, correcting for differences in conventions, in agreement
with \cite{AbbHar}.
Using (\ref{eqn:norm-soln},\ref{eqn:bog-mat-inf},\ref{eqn:bog-coeff}) in
(\ref{eqn:delta-defn}) one obtains, for waves re-entering the horizon in the
matter dominated era
\begin{equation}
\Delta = {H^2\over 2(2\pi)^3}\left[ {3j_1(k\tau)\over k\tau} \right]^2
\Rightarrow \Delta_{GW} = {H^2\over\pi^2 m_{Pl}^2}
\left[{3j_1(k\tau)\over k\tau}\right]^2 \delta_{\lambda\lambda'}
\end{equation}
where in the last step the scalar result was multiplied by $16\pi G$ to
obtain the corresponding result for gravitational waves.  The corresponding
expression for waves entering during the matter dominated era can be obtained
by replacing the factor in brackets by $j_0(k\tau)$ as one would expect
(since $j_0$ gives the right $\tau$ dependence from (\ref{eqn:wave-evol}) and
the amplitude as $k\tau\rightarrow 0$ should be the same).

We match this to our classical ensemble of gravitational waves,
$h_{\lambda}(\tau;\vec{k})$, by writing
\begin{equation}
  h_{\lambda}(\tau ;\vec{k}) = A(k) a_{\lambda}(\vec{k})
  \left[{3j_1(k\tau)\over k\tau}\right]\ \ \ \mbox{with}\ \ \lambda=+,\times
\label{eqn:spectral-amp}
\end{equation}
where the term in $[\cdots]$ is the real solution to (\ref{eqn:wave-evol}) in
the matter dominated phase and $a_{\lambda}(\vec{k})$ is a random variable
with statistical expectation value
\begin{equation}
\left\langle a_{\lambda}(\vec{k}) a_{\lambda'}(\vec{q}) \right\rangle
= k^{-3} \delta^{(3)}(\vec{k}-\vec{q}) \delta_{\lambda\lambda'}
\label{eqn:rv-norm}
\end{equation}

The 2-point function analogous to $\Delta$ in this case is simply
\begin{equation}
  \Delta'_{\lambda\lambda'} = A^2(k)\left[{3j_1(k\tau)\over k\tau}\right]^2
  \delta_{\lambda\lambda'}
\end{equation}
Matching the quantum and classical 2-point functions gives us the well known
prediction for the ($k$-independent) spectrum of gravitational waves generated
by inflation
\begin{equation}
A^2(k) = {H^2\over\pi^2 m_{Pl}^2} = {8\over 3\pi}v
\label{eqn:amplitude}
\end{equation}

The reason for the explicit factor of $k^3$ introduced in the definition of
$\Delta$ (\ref{eqn:delta-defn}) can be seen by considering the expression for
the spectral energy density in (classical) gravitational waves.
In our notation
\begin{equation}
  \rho = {k^2_{phys}\over 16\pi G}
    \left\langle h_{+}^2 + h_{\times}^2 \right\rangle
\end{equation}
where $\langle\cdots\rangle$ indicates an average over many wavelengths/periods
as well as an average over the stochastic variable $a_{\lambda}(\vec{k})$.
$k_{phys}=k/R(\tau)$ is the physical wavenumber.
For a fixed (non-stochastic) spectrum of waves we could write the average over
space as an integral over Fourier modes using Parseval's theorem and then
relate the ``power spectrum" $|h_{+,\times}(k)|^2$ to the correlation function
(\ref{eqn:delta-defn}).  With (\ref{eqn:spectral-amp}) and (\ref{eqn:rv-norm})
as defined the stochastic average produces the same result.  Explicitly
\begin{equation}
  k {d\rho_{\lambda}\over dk} = {k_{phys}^2\over 4G}
  \ \overline{\Delta'_{\lambda\lambda}}
  \ \ \ \ \ \ \lambda=+,\times
\end{equation}
which goes as $R^{-4}(\tau)$ as expected (an overbar denotes averaging).
Up to the time evolution factor $\overline{(3j_1(k\tau)/k\tau)^2}$ for matter
and $\overline{j_0(k\tau)}$ for radiation dominated phases this agrees with
\cite{KolTur}.
Including this factor for waves re-entering during the matter dominated phase
(i.e. considering evolution as the wave enters the horizon) leads to a
significant suppression of $\rho$.
Writing the energy density as a fraction of the closure density one finds, for
waves just entering the horizon \cite{KraWhi},
\begin{equation}
\Omega_g\equiv\sum_{\lambda=+,\times}{k\over\rho_c}
\left. {d\rho_{\lambda}\over dk} \right|_{k=2\pi/\tau} = \left\{
\begin{array}{cc} 16v/ 9 & \mbox{RD} \\ v/\pi^2 & \mbox{MD}\end{array}\right.
\end{equation}
In deriving this use has been made of the fact that $k_{phys}=1(2)\pi H_{HC}$
for waves crossing the horizon during the matter (radiation) dominated phase
($H_{HC}$ is the Hubble constant at the time the wave re-enters the horizon).

Will we be able to see this energy density with terrestrial or astrophysical
gravitational wave detector?
The most sensitive gravitational wave limit currently comes from the timing of
millisecond pulsars.  This is sensitive to ``short" wavelengths (periods of
order years) and puts an upper limit on the energy density in such waves
$\Omega_g<9\times 10^{-8}\ (68\%)$ (this limit improves as the fourth power
of the observing time: $T_{obs}^4$) \cite{pulsar}.
Such short wavelength modes entered the horizon during the radiation dominated
era.  If we write the Hubble constant today $H_0=100h$ km/s/Mpc then these
waves contribute to $\Omega_g$ today an amount suppressed by a factor of
$\rho_{rad}/\rho_c\approx 4\times 10^{-5}h^{-2}$ compared to horizon crossing
since they redshift with one extra power of $R$ compared to matter.
For values of $v\sim 10^{-9}$, consistent with the anisotropy of the CMBR (see
later), the energy density per logarithmic frequency interval in short
wavelength modes is predicted to be $\Omega_g\sim 10^{-13}$ \cite{KraWhi},
which is $\sim6$ orders of magnitude lower than the millisecond pulsar limit.
Even the proposed LIGO detector \cite{LIGO} which has a sensitivity of
$\Omega_g\sim 10^{-11}$ is short of the mark by 2 orders of magnitude.
Thus we expect that a positive detection of the signal will require a
significant advance in technology.

Since the predicted CMBR anisotropy multipoles, up to $l\sim 9$, from
gravitational waves and scale invariant scalar fluctuations are very similar
\cite{AbbSch} and direct detection of gravitational waves is still some way off
the outlook for determining unambiguously that a significant fraction of the
observed CMBR anisotropy comes from gravitational waves seems bleak.

\vspace{.1in}

\noindent 5. {\bf Anisotropy from the stochastic background}

It is conventional to expand the CMBR temperature anisotropy in spherical
harmonics
\begin{equation}
  {\delta T\over T}(\theta,\phi) = \sum_{lm} a_{lm} Y_{lm}(\theta,\phi)
\label{eqn:mode-expansion}
\end{equation}
We can calculate the prediction of a given spectrum of gravitational waves in
terms of the $a_{lm}$.
The temperature fluctuation due to a gravitational wave $h_{\mu\nu}$ can be
found in the linearized theory to be \cite{SacWol}
\begin{equation}
  {\delta T\over T} = -{1\over 2} \int_e^r d\Lambda
  \ {\partial h_{\mu\nu}(\tau,\vec{x})\over \partial\tau}
  \hat{x}^{\mu} \hat{x}^{\nu}
\label{eqn:sachs-wolfe}
\end{equation}
where $\Lambda$ is a parameter along the unperturbed path and the lower (upper)
limit of integration represents the point of emission (reception) of the
photon.

Now we project out a multipole and calculate the rotationally symmetric
quantity
\begin{equation}
  \left\langle a_l^2 \right\rangle \equiv
  \left\langle \sum_m |a_{lm}|^2 \right\rangle =
  \sum_m \int d\Omega d\Omega'\ Y_{lm}^{*}(\Omega)Y_{lm}(\Omega')
  \left\langle{\delta T\over T}(\Omega){\delta T\over T}(\Omega')\right\rangle
\end{equation}
where using (\ref{eqn:plane-wave},\ref{eqn:rv-norm},\ref{eqn:sachs-wolfe})
we have for a spectrum of waves
\begin{eqnarray}
 \left\langle{\delta T\over T}(\Omega){\delta T\over T}(\Omega')\right\rangle
 &=& {1\over 4}\sum_{\lambda=+,\times}\int {d^3k\over k^3}
\left[ \int d\Lambda\ \dot{h}(\tau ; k) e^{i\vec{k}\cdot\vec{x}}
  \epsilon_{\mu\nu}(\vec{k};\lambda) \hat{x}^{\mu}\hat{x}^{\nu} \right]
\times \nonumber \\
& &\left[ \int d\Lambda'\ \dot{h}(\tau' ; k) e^{i\vec{k}\cdot\vec{x'}}
  \epsilon_{\mu\nu}(\vec{k};\lambda) \hat{x'}^{\mu}\hat{x'}^{\nu} \right]^{*}
\end{eqnarray}
and an overdot represents a derivative with respect to $\tau$.
We use the rotational symmetry of $a_l^2$ to make the replacement
\begin{equation}
  \sum_m Y_{lm}^{*}(\Omega) Y_{lm}(\Omega')
= \sum_m Y_{lm}^{*}(\Omega_{kx}) Y_{lm}(\Omega'_{kx})
\end{equation}
where $\Omega_{kx}$ indicates that the angles are defined with respect to
$\hat{k}$.  We can evaluate directly
\begin{equation}
\epsilon_{\mu\nu}(\vec{k};\lambda) \hat{x}^{\mu}\hat{x}^{\nu} =
\left( \delta_{\lambda}^{+}\cos(2\phi) + \delta_{\lambda}^{\times}\sin(2\phi)
  \right)\sin^2\theta
\end{equation}
where $\theta, \phi$ are the usual spherical angles relating $\hat{k}$ and
$\hat{x}$.

First concentrate on the angular integrations.  Expanding the exponential
\begin{equation}
  e^{i\vec{k}\cdot\vec{x}} = \sum_{n=0}^{\infty}
       i^n (2n+1) j_n(kx) P_n(\cos\theta_{kx})
\end{equation}
it is not hard to show (formula 7.125 of Gradshteyn and Ryzhik \cite{GraRyz}
is useful) that
\begin{eqnarray}
I_{lm}(k,x) & \equiv &
\int d\Omega_{kx}\ Y_{lm}^{*}(\Omega_{kx})\ e^{i\vec{k}\cdot\vec{x}}
  \epsilon_{\mu\nu}(\vec{k};\lambda) \hat{x}^{\mu} \hat{x}^{\nu} \nonumber \\
  &=& \pi\sqrt{{2l+1\over 4\pi} {(l+2)!\over (l-2)!}}\  {\cal H}
  \sum_{n} i^n (2n+1) j_n(kx)
  \left( c_{-2}\delta_{l-2}^n + c_0\delta_l^n + c_2 \delta_{l+2}^n \right)
\end{eqnarray}
where
\begin{eqnarray}
{\cal H} & = & (\delta_{\lambda}^{+}-i\delta_{\lambda}^{\times})\delta_m^{+2}
  + (\delta_{\lambda}^{+}+i\delta_{\lambda}^{\times})\delta_m^{-2} \\
1/c_{-2} & = & (2l-1)(2l+1)(l-3/2) \\
1/c_{ 0} & = &-{1\over 2}(2l-1)(2l+3)(l+1/2) \\
1/c_{+2} & = & (2l+1)(2l+3)(l+5/2)
\end{eqnarray}

Inserting this back into the expression for $a_l^2$ and using
(\ref{eqn:spectral-amp})
\begin{equation}
\left\langle a_l^2 \right\rangle = {1\over 4}\sum_m\sum_{\lambda=+,\times}
\int {dkd\Omega_k\over k} \left| \int d\Lambda\ A(k) {d\over d\tau}
\left({3j_1(k\tau)\over k\tau}\right) I_{lm}(k,x) \right|^2
\end{equation}

The integral over the path $d\Lambda$ can be parameterized by the distance
from the origin along the line of sight, $r$, so $|\vec{x}(r)|=r$ and
$\tau(r)=\tau_0-r$.  We will defer consideration of the limits of the $k$
integral for the moment.  Thus
\begin{equation}
\left\langle a_l^2 \right\rangle = 36\pi^2 (2l+1){(l+2)!\over (l-2)!}
\int k dk\ A^2(k) | F_l(k) |^2
\label{eqn:moment}
\end{equation}
where the function $F_l(k)$ is defined as
\begin{equation}
F_l(k) \equiv
\int_0^{\tau_0-\tau_1} dr \left({d\over d(k\tau)}{j_1(k\tau)\over k\tau}\right)
\left[ {j_{l-2}(kr)\over(2l-1)(2l+1)}+{2j_l(kr)\over(2l-1)(2l+3)}
      +{j_{l+2}(kr)\over(2l+1)(2l+3)} \right]
\label{eqn:flk-defn}
\end{equation}
Accounting for the factor of two difference between the definitions of $A^2(k)$
this is precisely the result of \cite{AbbWis1}, and is a factor of
$\approx2$ larger than the earlier result of \cite{FabPol}.  The integral in
equation (\ref{eqn:flk-defn}) can be expressed in terms of elementary
functions and the sine integral but the result is very cumbersome and is not
presented here.

For waves entering the horizon during the matter dominated regime the upper
limit of the $k$ integration is $2\pi/\tau_1$.  The results are very
insensitive to the exact choice of upper limit since the integral is
dominated by small $k\approx 2\pi /\tau_0$, i.e. waves that have recently
entered the horizon.
For very high frequency modes the transition between phases of the universe
is no longer sudden and graviton production is suppressed.
Taking this high frequency cutoff to be $k=2\pi/\tau_1$ (i.e. restricting
attention to waves that entered the horizon during the matter dominated era)
introduces negligible error.
The lower limit of the $k$ integral is zero, since waves of arbitrarily long
wavelength can contribute\footnote{We thank Mark Wise and Vince Moncrief for
very helpful discussions on this point.}.  From the form of the integrand
however one can see that the contribution to $\langle a_l^2\rangle$ tends to
zero as $k^{n}\ (n\ge 1)$ as $k\tau\rightarrow 0$.  Thus the contribution
from {\em very} long wavelength modes is suppressed as one might expect
on physical grounds.

One can now calculate the predicted $\left\langle a_l^2 \right\rangle$ for any
spectrum $A(k)$ of gravitational waves from (\ref{eqn:moment}).
For the scale-invariant spectrum (\ref{eqn:amplitude}) of exponential inflation
the $\langle a_l^2 \rangle$ are shown in table \ref{tab:moments}.
The calculation of the expectation value $\langle a_l^2 \rangle$ is not the end
of the story however.  Before we compare predictions with observations we must
also consider the statistical properties of $a_l^2$.  The fact that the
``weakly coupled" nature of inflation predicts the $a_{lm}$ to be independent
makes this problem tractable.  Given that each of the $a_{lm}$ are independent
Gaussian random variables the probability distribution for each $a_l^2$, whose
mean $\langle a_l^2\rangle$ we calculated, is of the $\chi^2$ form and can be
written
\begin{equation}
  P(y) dy = {y^{l-1/2} e^{-y}\over \Gamma\left( l+{1\over 2} \right)}  dy
\ \ \ \ \mbox{ where }\ \ \ y = {2l+1\over 2}{a_l^2\over \langle a_l^2\rangle}
\label{eqn:chi-square}
\end{equation}
This agrees with \cite{AbbWis2}. One can calculate the confidence
levels for $a_l^2$ in terms of the incomplete gamma function (values for the
68, 90 and 95\% (lower) confidence levels can be found in table
\ref{tab:moments}).
We note in passing that the modal value of $a_l^2$ for any universe is
$(2l-1)/(2l+1)$ times the mean.

\vspace{.1in}

\noindent 6. {\bf The correlation function}

Limits quoted on the multipole moments $a_l^2$ can be directly compared to
(\ref{eqn:moment}) to constrain the inflationary theory.  However many
experiments (and importantly COBE) are of the double- or triple-beam type and
do not simply measure any one multipole moment but can constrain the
correlation function \cite{COBE,KolTur}
\begin{equation}
  C(\theta_{21};\sigma) \equiv \left\langle {\delta T\over T}(\hat{x}_1;\sigma)
  {\delta T\over T}(\hat{x}_2;\sigma) \right\rangle_{21}
\label{eqn:corr-defn}
\end{equation}
where the average is over all positions $\hat{x}_1,\hat{x}_2$ on the sky with
$\hat{x}_1\cdot\hat{x}_2$ fixed, and $\sigma$ is a measure of the angular
response of the detector.  We shall take the angular response to be Gaussian
\cite{WilSil}
\begin{equation}
  dR(\theta,\phi) = {\theta d\theta d\phi\over 2\pi\sigma^2}
  \exp\left[ -{\theta^2\over 2\sigma^2} \right]
\end{equation}
where $\theta,\phi$ are the angles relative to the beam direction and
$\sigma\ll 1$ which justifies the use of the small angle expansion.  We define
the ``smeared" $\delta T/T$ by
\begin{eqnarray}
{\delta T\over T}(\hat{x}_1;\sigma) &\equiv&
      \sum_{lm} a_{lm} \int d\Omega'\ R(\Omega') Y_{lm}(\Omega) \\
& = & \sum_{lm} a_{lm} \left( \int_0^{\infty} dx\ P_l(1-\sigma^2x)e^{-x}\right)
      Y_{lm}(\hat{x}_1)
\end{eqnarray}
where $\Omega'$ is the direction $\Omega$ measured relative to $\hat{x}_1$ and
we have approximated $\cos\theta'\approx 1-\theta'^2/2$ for small $\theta'$.
Note that the term in parenthesis suppresses the higher $l$ modes as we would
expect physically.  We can expand this term for small $\sigma^2$
\begin{eqnarray}
\int_0^{\infty} dx\ P_l(1-\sigma^2x)e^{-x}
&=& \sum_{n=0}^{\infty} (-\sigma^2)^n P_l^{(n)}(1) \\
&=& \sum_{n=0}^{l} (-\sigma^2)^n {1\over 2^n n!} {(l+n)!\over (l-n)!}
\end{eqnarray}
Making the approximation $(l+n)!/(l-n)!\approx (l+1/2)^{2n}$ for large $l$ we
can write the suppression term as: $\exp\left[-(l+1/2)^2\sigma^2/2\right]$ (and
at the same level of approximation we can replace $(l+1/2)^2$ with $l(l+1)$ as
used by \cite{COBE}).
This exponential suppression proves a good approximation for $\sigma<0.1$
which is the range for most detectors.
A simple exercise in manipulating spherical tensors allows us to write
\begin{equation}
\left\langle Y_{lm}(\hat{x}_1) Y_{l'm'}^{*}(\hat{x}_2)\right\rangle_{21}
= {1\over 4\pi}\ \delta_{ll'}\ \delta_{mm'}\ P_l(\hat{x}_1\cdot\hat{x}_2)
\end{equation}
so we can express the correlation function (\ref{eqn:corr-defn}) in a simple
from as (see \cite{COBE,KolTur})
\begin{equation}
C\left(\hat{x}_1\cdot\hat{x}_2;\sigma\right)={1\over 4\pi} \sum_{l=0}^{\infty}
a_l^2 P_l\left(\hat{x}_1\cdot\hat{x}_2\right) e^{-(l+1/2)^2\sigma^2}
\end{equation}

The predicted correlation function using the moments of table
\ref{tab:moments} is shown in figure \ref{fig:correlation} for a gaussian beam
width of $\sigma=10^{o}$.
The error bars represent the upper and lower limit of the 68\% confidence
region arrived at using (\ref{eqn:chi-square}) for the distribution of
$a_l^2/v$ (there is no simple analytic form for the probability distribution
of $C(\theta,\sigma)$ so figure \ref{fig:correlation} was obtained from a
Monte-Carlo) and the solid line is $C(\theta,\sigma)$ evaluated using
$a_l^2=\langle a_l^2 \rangle$.
The correlation function inferred from the COBE sky maps \cite{COBE} is in
agreement with the form shown in figure \ref{fig:correlation}.

\vspace{.1in}

\noindent 7. {\bf Comparison with observations}

The recent results of COBE can be summarized as a non-zero value of the
quadrupole moment $a_2^2$.  If we require a Harrison-Zel'dovich spectrum the
quadrupole, from the $Q_{rms-PS}$ value, is measured to be \cite{KraWhi,COBE}
\begin{equation}
   a_2^2 = \left( 4.7\pm 2\right) \times 10^{-10}
\end{equation}

An early inflationary phase produces not only gravitational waves but also
a scale invariant spectrum of scalar density perturbations.  To limit the
contribution of these latter to $a_2^2$ we can require the induced dipole from
long wavelength scalar modes not greatly exceed the observed dipole anisotropy.
For a given (e.g. flat) spectrum one can use this to place limits on all the
higher multipoles.  At the 90\% confidence level an upper limit of
$a_2^2\approx 2\times 10^{-10}$ \cite{AbbSch} has been derived (ignoring the
transfer function corrections).
Equivalently fitting the observed clustering to a primordial spectrum yields
$a_2^2\approx (2-10)\times 10^{-11}$ \cite{BonEfs}.  Such estimates suggest a
large fraction of the quadrupole anisotropy may be due to long wavelength
gravitational waves.

Let us assume that the observed quadrupole is due entirely to gravitational
waves.
Using $\langle a_2^2\rangle /v$ from table \ref{tab:moments} and including
the distribution for the measured $a_2^2$ (gaussian) and predicted $a_2^2/v$
(\ref{eqn:chi-square}) one can infer a distribution for $v$.
The result obtained from a Monte-Carlo is shown in figure
\ref{fig:monte-carlo}.
This corresponds to a mean $v=6.1\times 10^{-11}$, and a modal
$v=4\times 10^{-11}$ with non-gaussian errors.  The allowed range of $v$ from
figure \ref{fig:monte-carlo} can be summarized as \cite{KraWhi}
\begin{equation}
\begin{array}{cccccc}
 2.3 \times 10^{-11} & < & v & < & 1.5 \times 10^{-10} & 68\%\  CL \\
 2.5 \times 10^{-12} & < & v & < & 3.7 \times 10^{-10} & 95\%\  CL
\end{array}
\label{eqn:limits}
\end{equation}
One can scale these results directly to obtain values for $v$ if gravitational
wave contribution to $a_2^2$ is not 100\%, the rest being made up or shielded
by scalar fluctuations for example.

Converting to a more familiar energy scale $M=(v^{1/4}M_{Pl})$ the 95\%
confidence level range above is $M\in (1.5-5.2)\times 10^{16}$GeV, with a best
fit $M=3\times 10^{16}$GeV.
The upper limit quoted corresponds to a strict upper limit on the scale of
inflation assuming that the quadrupole moment due to gravitational waves is
not being cancelled to any significant degree by other sources (e.g. scalar
density perturbations).
A large cancellation between two sources of anisotropy would be unlikely.

No realistic model of inflation currently exists, however most are based on
the idea that a phase transition occurred in the early universe to produce the
vacuum energy density to drive the inflation.
If such a transition were to occur then the natural scale would seem to be the
scale of unification of the coupling constants in the context of a Grand
Unified Theory (GUT) (an exception is the case of chaotic inflation).
The recent precision measurements from LEP have allowed us to sharpen our
predictions for this scale.
Assuming a supersymmetric GUT with the supersymmetry breaking scale
$\approx 1$TeV, one predicts \cite{couplings} a unification scale of
$(1-3.6)\times 10^{16}$GeV.
For a theory based on SO(10) one finds $M = 10^{15.8\pm .22}$GeV \cite{MohPar}.
The limits from proton decay experiments are very theory dependent, however
they give a {\em lower} bound on $M$.  The current limits are near or larger
than $10^{15}$GeV for fashionable GUT theories.

Unless there is significant fine tuning or hierarchies in the theory the
vacuum energy density associated with a transition from a GUT with scale $M_G$
is $V_0=(\kappa M_G)^4$ where $\kappa$ is a (calculable) number usually in the
range $\kappa^4=.01-1$.  This puts the GUT scale very close to the scale
of inflation inferred above -- a ``numerical coincidence" which is both
suggestive and exciting.

\vspace{.1in}

\noindent 8. {\bf Conclusions}

As the observations of CMBR anisotropies are refined and other gravitational
wave detectors come on line the results presented here can allow a comparison
of theoretical predictions and experimental results, which can help to
constrain models of inflation and early universe physics.

The recent COBE results allow the possibility that the observed quadrupole
anisotropy results from gravitational waves generated during inflation with a
scale of $(1.5-5.2)\times 10^{16}$GeV -- a very interesting range of energies.
If the anisotropy {\em is} due to gravitational waves, confirmation will
probably not come from existing or currently planned gravitational wave
detectors.  Further results are eagerly awaited.

\vspace{.3in}

I would like to thank L. Krauss, V. Moncrief, M. Turner and M. Wise for
extremely useful conversations.  I also wish to thank E. Gates for comments
on the manuscript.

\clearpage

\centerline{ {\bf Figure Captions} }

\begin{figure}[h]
\caption{Results of a Monte-Carlo for the predicted correlation function
$C(\theta;\sigma)$ generated by a flat spectrum of gravitational waves.
The error bars represent the upper and lower limits of the 68\% confidence
region arrived at using the predicted distribution of $a_l^2/v$. The solid
curve is the correlation function calculated using $\langle a_l^2 \rangle$.
The beam width is taken to be $\sigma=10^{o}$.}
\label{fig:correlation}
\end{figure}

\begin{figure}[h]
\caption{The probability distribution for the scale $v\equiv V_0/m_{Pl}^4$ of
inflation as determined by Monte-Carlo from the COBE measurement of the
quadrupole anistropy assuming that the whole anistropy is due to gravitational
waves.}
\label{fig:monte-carlo}
\end{figure}

\vspace{.5in}

\centerline{ {\bf Tables} }

\begin{table}[h]
\begin{center}
\begin{tabular}{|c|c|c|}
\hline
      &        &               \\
      &        & $a_l^2/\langle a_l^2\rangle$ \\
      &        & \\
$l$   & $\langle a_l^2\rangle/v$ & 68/90/95\% CL \\ \hline
 2    &  7.74  & .63/.32/.23   \\
 3    &  4.25  & .69/.40/.31   \\
 4    &  3.10  & .73/.46/.37   \\
 5    &  2.50  & .76/.51/.42   \\
 6    &  2.12  & .78/.54/.45   \\
 7    &  1.85  & .80/.57/.48   \\
 8    &  1.64  & .81/.59/.51   \\
 9    &  1.48  & .82/.61/.53   \\
 10   &  1.35  & .83/.63/.55   \\ \hline
\end{tabular}
\end{center}
\caption{Multipole coefficients $a_l^2$ for modes $l=2-10$ predicted for a
stochastic background of gravitational waves generated by exponential
inflation.  The 68, 90 and 95\% (lower) confidence levels are also shown as
fractions of $\langle a_l^2 \rangle$.}
\label{tab:moments}
\end{table}

\end{document}